\documentclass{PoS}

\input{babarsym}
\title{$\babar$: sin($2\beta+\gamma$)}

\ShortTitle{$\babar$: sin($2\beta+\gamma$)}

\author{\speaker{Cecilia Voena}\thanks{On behalf of the $\babar$ collaboration}\\\
        Universit\`a ``La Sapienza'' di Roma and INFN Roma, Italy\\
        E-mail: \email{cecilia@slac.stanford.edu}}

\abstract{The time-dependent $CP$ asymmetries in fully reconstructed 
$\Bz{\to}D^{(*)\pm}\pi^{\mp}/\rho^{\mp}$ decays (new preliminary result), and in partially reconstructed  $\Bz{\to}D^{(*)\pm}\pi^{\mp}$ decays, are measured with the 
$\babar$ detector at the PEP-II asymmetric $B$ factory at SLAC, 
using $232$ million $\FourS \rightarrow \BB$ decays. 
We combine the above results and, using other measurements and theoretical assumptions, we interpret them in terms of the angles of the unitarity triangle describing the Cabibbo-Kobayashi-Maskawa matrix.  We find $|\rm{sin}(2\beta+\gamma)|>0.64(0.42)$ at $68\%$($90\%$) confidence level using a frequentistic approach and $|2\beta+\gamma|$=$(90\pm43)^o$ using a Bayesian approach.}

\FullConference{International Europhysics Conference on High Energy Physics\\
                 July 21st - 27th 2005\\
                 Lisboa, Portugal}

\begin{document}

\section{Introduction}
In the Standard Model, $CP$ violation in the weak interactions between quarks
manifests itself as a non-zero area of the Cabibbo-Kobayashi-Maskawa (CKM) unitarity triangle~\cite{CKM}.
Constraints on the quantity $\rm{sin}(2\beta+\gamma)$ can be obtained from the 
study of the time evolution of $\Bz{\to} D^{(*)\pm} \pi^{\mp}$ and $\Bz{\to} D^{\pm} \rho^{\mp}$ decays~\cite{sin2bg} that is sensitive to $\gamma$ due to interference between the CKM-favored $b\rightarrow c$ and the CKM-suppressed $b\rightarrow u$ transitions through $\BzBzb$ mixing.

The $\babar$ detector~\cite{nim} collects data at the PEP-II asymmetric $e^+e^-$ collider operated at or near the $\FourS$ resonance. $\BB$ pairs from the $\FourS$ decay move along the high-energy beam direction with a Lorentz boost $\beta\gamma$=0.56. The time dependent distribution for $\Bz$ decays to a final state $\mu$ ($\mu=D\pi$, $D^*\pi$ and $D\rho$), is given by (neglecting the decay width difference):
\begin{eqnarray}
f^{\pm,{\mu}}(\eta,\Delta t) &=&
 \frac{e^{-\left|\Delta t\right|/\tau}}{4\tau} \times [ 1 \mp (a^{\mu}
  \mp \eta b - \eta c^{\mu})\nonumber \\
  && \sin(\Delta m_d \Delta t)\mp\eta\cos(\Delta m_d \Delta t)]\,,
\label{eq}
\end{eqnarray}
where $\tau$ is the $\Bz$ lifetime, $\Delta m_d$ is the $\BzBzb$ mixing frequency and $\Delta t = t_{\rm rec} - t_{\rm tag}$ is the time of the
$\Bz{\to} D^{(*)\pm} \pi^{\mp}$ or $\Bz{\to} D^{\pm} \rho^{\mp}$ decay ($B_{\rm rec}$) relative to the decay of the other $B$ ($B_{\rm tag}$) from the $\FourS \rightarrow \BB$ decay. $\Delta t$ is calculated from the measured  separation along the
beam collision axis ($z$), between the $B_{\rm rec}$ and $B_{\rm tag}$ decay vertices: $\Delta z$=$\beta\gamma c \Delta t$. In equation~\ref{eq} the upper (lower) sign refers to the flavor of $B_{\rm tag}$ as $\Bz$ $(\Bzb)$,
while $\eta=+1$ ($-1$) for the final state with a $D^{(*)-}$ ($D^{(*)+}$).
In the Standard Model:
\begin{eqnarray}\nonumber
&a^{\mu}&=\ 2r^{\mu}\sin(2 \beta+\gamma)\cos\delta^{\mu}\,, \\ \nonumber
&b&=\ 2r^\prime\sin(2 \beta+\gamma)\cos\delta^\prime\,, \\
&c^{\mu}&=\ 2\cos(2 \beta+\gamma) (r^{\mu}\sin\delta^{\mu}-r^\prime\sin\delta^\prime)\,.
\label{acdep}
\end{eqnarray}
 The parameters $r^{\prime}$ and $\delta^{\prime}$ accounts for non-negligible $CP$ violating effects on the $B_{tag}$ side that depend on the type of events used to determine the $B_{tag}$ flavor. We use only the $c^{\mu}$ parameters of events tagged with leptons ($c^{\mu}_{\rm{lep}}$) and the $a^{\mu}$ parameters in the determination of $\rm{sin}(2\beta+\gamma)$ since they are independent of $r^{\prime}$ and $\delta^{\prime}$ (semileptonic $B$ decays have $r^{\prime}_{\rm lep}=0$).

In the following the analysis based on fully reconstructed $\Bz\rightarrow D^{(*)\pm}\pi^{\mp}/\rho^{\mp}$ decays (Section~\ref{fully}), and with partially reconstructed $\Bz\rightarrow D^{(*)\pm}\pi^{\mp}$ decays (Section~\ref{partial}) are discussed. Section~\ref{interpret} presents the interpretation of the results in terms of $\rm{sin}(2\beta + \gamma)$.

\section{Fully reconstructed $\Bz\rightarrow D ^{(*)\pm}\pi^{\mp}/\rho^{\mp}$ decays}\label{fully}
The analysis is based on a sample of 15635 $\Bz\rightarrow D^{\pm}\pi^{\mp}$, 14554 $\Bz \rightarrow D^{*\pm}\pi^{\mp}$ and 8736 $\Bz\rightarrow D^{\pm}\rho^{\mp}$ decays selected from $232$ million $\FourS$ $\rightarrow$ $\BB$ decays~\cite{fully}. The purity of the samples is $82\%-93\%$ depending on the decay mode.
We reconstruct the $B_{\rm rec}$ by fully reconstructing all the particles in the final state. $\Bz$ candidates are selected using the difference between the energy of the candidate and the beam energy $\sqrt{s}/2$ in the center of mass frame, and the beam energy substituted mass, calculated from $\sqrt{s}/2$ and the reconstructed momentum of the $\Bz$.
After the selection, two kinds of backgrounds remain: combinatorial background from random combination of tracks in the events, estimated from data, and background coming from misreconstructed $B$ decays, that may peak in the energy substituted mass near the $B$ mass and is determined from simulation.

 We determine the $B_{\rm rec}$ decay vertex from its charged tracks.
The  $B_{\rm tag}$ decay vertex is obtained by fitting tracks that do not belong to $B_{\rm rec}$ using constraints from the known beam energy and from the position of the luminous region.
The $\Delta t$ resolution is approximately $1.1$ ps.

We use multivariate algorithms that identify signatures in the $B$ decay products that determine the flavor of the $B_{tag}$ to be either a $\Bz$ or a $\Bzb$ as primary leptons from semi-leptonic $B$ decays, kaons and soft pions from $D^{*}$ decays. We combine them using a neural network and we assign each event with mistag probability less than $45\%$ to one of six hierarchical, mutually exclusive tagging categories. The effective efficiency of the tagging algorithm, defined as $Q = \Sigma_i \epsilon_i(1-2w_i)^2$, where $\epsilon_i$ and $w_i$ are the efficiency and the mistag probability for the tagging category $i$, is $(30.5\pm0.4)\%$.

\begin{figure}[!htb]
\begin{center}
\includegraphics[height=9cm]{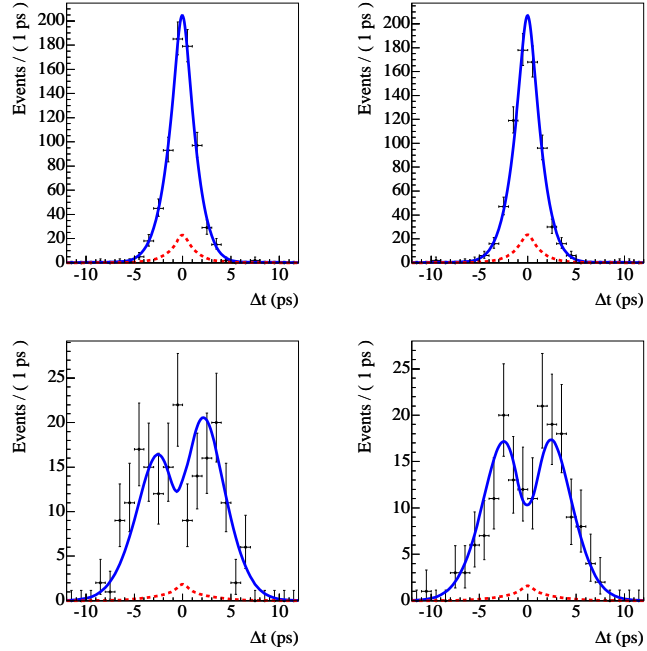}
\caption{$\Delta t$ distribution on data for the decay mode
$\Bz \rightarrow D^{*\pm}\pi^{\mp}$ in the lepton tagging category for, from upper left going clockwise, $B_{tag}=\Bzb$ and $B_{rec}=D^{*-}\pi^+$, $B_{tag}=\Bz$ and $B_{rec}=D^{*+}\pi^-$, $B_{tag}=\Bz$ and $B_{rec}=D^{*-}\pi^+$, $B_{tag}=\Bzb$ and $B_{rec}=D^{*+}\pi^-$. The result of the fit is superimposed. In each plot, the dashed curve
represents the background contribution. }\label{mes}
\end{center}
\end{figure}

An unbinned maximum-likelihood fit is performed  to the  $\Delta t$ distribution of Eq.~\ref{eq}, appropriately modified to take into account the finite detector $\Delta t$ resolution, the mistag probability, and every significant source of background. 
The resolution function and the mistag parameters as well as many background parameters are determined from the fit to the data simultaneously with the $CP$ parameters. From the unbinned maximum likelihood fit we obtain: $a^{D\pi} = -0.013\pm0.022 \,(\textrm{stat.})\pm 0.007 \,(\textrm{syst.})$, $c_{\rm lep}^{D\pi} = -0.043\pm0.042  \,(\textrm{stat.})\pm 0.011 \,(\textrm{syst.})$, $a^{D^*\pi} = -0.043\pm0.023  \,(\textrm{stat.})\pm 0.010 \, \\(\textrm{syst.})$, $c_{\rm lep}^{D^*\pi}  = 0.047\pm0.042  \,(\textrm{stat.})\pm 0.015 \, (\textrm{syst.})$, $a^{D\rho} = -0.024 \pm0.031  \,(\textrm{stat.})\pm 0.010 \,(\textrm{syst.})$ and  $c_{\rm lep}^{D\rho}  = -0.098\pm0.055  \,(\textrm{stat.})\pm 0.019 \,(\textrm{syst.})$.
Figure~\ref{mes} shows the $\Delta t$ distribution for events tagged with leptons  for the sample with the highest purity, $\Bz \rightarrow D^{*\pm}\pi^{\mp}$.
The biggest contribution to the systematic uncertainty comes from effects releated to the measurement of $\Delta t$.

\section{Partially reconstructed $\Bz\rightarrow D^{(*)\pm}\pi^{\mp}$ decays}\label{partial}
The analysis is based on 89290 $\Bz\rightarrow D^{*\pm}\pi^{\mp}$ decays selected from $232$ million $\FourS \rightarrow \BB$ decays~\cite{partial}. The $B$ are partially reconstructed using the  hard pion from the $B$ decay and the soft pion from the $D^*$ applying kinematic constraints consistent with the signal decay mode. Signal and background are obtained from fits to the computed $D$ mass on several data samples.
We determine the $B_{rec}$ decay vertex by fitting the hard pion track and the $B_{tag}$ decay vertex from a fit of all other tracks in the event excluding all tracks within 1 rad of the $D$ momentum in the center of mass frame.
The flavor of the $B_{tag}$ is determined using leptons or kaons.
The purity of the sample is $54\%$ and $31\%$ for events tagged with leptons and kaons respectively.
The $CP$ parameters are obtained with a unbinned maximum likelihood fit and the result is: $a^{D^*\pi} = -0.034\pm0.014  \,(\textrm{stat.})\pm 0.009 \,(\textrm{syst.})$, $c_{\rm lep}^{D^*\pi}  = -0.025\pm0.020  \,(\textrm{stat.})\pm 0.013 \,(\textrm{syst.})$.
The dominant systematic uncertainty comes from possible $CP$ violating contributions in the backgrounds.
\section{Summary and interpretation of the result}\label{interpret}
We have measured the time-dependent distribution of $\Bz \rightarrow D^{*\pm}\pi^{\mp}$ decays using both a fully reconstructed and a partially reconstructed sample and of $\Bz \rightarrow D^{\pm}\rho^{\mp}$ decays using a fully reconstructed sample, based on 232 million $B\bar{B}$ pairs.
We combine the above measurements to obtain constraints on $\rm{sin}(2\beta+\gamma)$ using both a frequentistic and a Bayesian approach~\cite{fitters}.
Given the small size of the $r_{D\pi}$, $ r_{D^*\pi}$ and $r_{D\rho}$ parameters ($\sim$0.02) we are not able to extract them from the fit to the data. Instead we estimate these parameters using $SU(3)$ symmetry relations~\cite{partial} and we assign a 30$\%$ (frequentistic approach) or 100$\%$ (Bayesian approach) uncertainty to $SU(3)$ breaking effects and other theoretical uncertainties. We find:
$|\rm{sin}(2\beta+\gamma)|>0.64(0.42)$ at $68\%$($90\%$) confidence level and $|2\beta+\gamma|$=$(90\pm43)^o$ in the frequentistic and in the Bayesian approach, respectively.

\end{document}